\documentclass[namedreferences]{solarphysics}
\usepackage[optionalrh]{spr-sola-addons} 
\usepackage{graphicx}        
\usepackage{color}           
\usepackage{url}             
\pdfoutput=1

\newcommand{\etal}{{\it et al.}}

\newcommand{\aap}{    {\it Astron. Astrophys.}}

\newcommand{\apj}{    {\it Astrophys. J.}}

\newcommand{\jgr}{    {\it J. Geophys. Res.}}
\newcommand{\mnras}{  {\it Mon. Not. Roy. Astron. Soc.}}

\newcommand{\solphys}{{\it Solar Phys.}}

\newcommand{\ssr}{    {\it Space Sci. Rev.}}

\begin{document}

\begin{article}

\begin{opening}

\title{{3-D nonlinear force-free field reconstruction of solar active region 11158 by direct boundary integral equation}}

\author{Rui~\surname{Wang}$^{1}$\sep
        Yihua~\surname{Yan}$^{1}$\sep
        Baolin~\surname{Tan}$^{1}$
       }
\runningauthor{Wang, Yan, \& Tan} \runningtitle{GPU-accelerated
DBIE NLFFF extrapolation}

\institute{$^{1}$ Key Laboratory of Solar Activity, National
Astronomical Observatories, Chinese Academy of Sciences, Beijing,
100012, China. email: \url{Ray@nao.cas.cn},\url{yyh@nao.cas.cn},\url{bltan@nao.cas.cn}}

\begin{abstract}

A 3-D coronal magnetic field is reconstructed for the NOAA active region
11158 on February 14, 2011. A GPU-accelerated direct boundary integral
equation (DBIE) method is implemented. This is approximately 1000 times faster than the original DBIE used on
solar non-linear force-free field modeling. Using the SDO/HMI vector magnetogram
as the bottom boundary condition, the reconstructed magnetic field lines are compared
with the projected EUV loop structures as observed by SDO/AIA at front view and the
STEREO A/B spacecraft at side views for the first time. They show very good agreement
three-dimensionally so that the topology configurations of the magnetic fields can be analyzed,
thus its role in the flare process of the active region can be better understood.  {A quantitative comparison with some stereoscopically reconstructed coronal loops shows that the present averaged
misalignment angles are at the same order
as the state-of-the-art results obtained with reconstructed coronal loops as prescribed conditions}. It is found that
the observed coronal loop structures can be grouped into a number of closed and open field
structures with some central bright coronal loop features around the polarity inversion line.
{The reconstructed
highly-shearing magnetic field lines agree very well with the low-lying sigmoidal filament along the polarity inversion line.
They are in a pivot position to all other surrounding coronal structures, and a group of electric current lines co-aligned
with the central bright EUV loops overlying the filament channel is also obtained. This central lower-lying magnetic
field loop system must have played a key role in powering the flare. It should be noted that while a strand-like coronal feature
along the polarity inversion line may be related to the filament, one cannot simply attribute all the coronal bright features
along the polarity inversion line to manifestation of the filament without any stereoscopical information.} The numerical
procedure and the comparison against a benchmark test case are also presented to validate that the DBIE method is rigorous
and effective.

\end{abstract}

\end{opening}

\section{Introduction}
     \label{Introduction}

It is well-known that the magnetic field plays a key role in almost all solar
activities, such as solar flares, filament eruptions, coronal mass ejections, etc. {Many
structures of the solar corona are shaped by the magnetic field, due to its pervasive nature.} Therefore, a thorough
knowledge about the coronal magnetic field topology will help
us to understand the physical processes of various
solar activities. {However, so far the routine measurement of
solar magnetic field is mainly based on the Zeeman effect and Hanle effect, which
can measure stronger emissions and sharper lines on the photosphere but
failed to measure the coronal magnetic field for its faint line intensities and broad line widths.} Although some
techniques using infrared and radio observations have been
proposed to solve this problem, (Lin et al., 2004; Gary and Hurford, 1994),
they have not reached full fruition yet and have many limitations. Normally, one has to obtain the
solar coronal magnetic fields from modeling by extrapolation using underlying photospherical observations.

At present, the non-linear force-free field (NLFFF) model has been thought
to be a good approximation to the actual physical state of the
coronal magnetic fields. Available NLFFF extrapolation methods can be classified into five types:
(1) Upward integration
method: Nakagawa, 1974; Wu et al., 1990; Song et al., 2006; (2)
Grad-Rubin method: Grad and Rubin, 1958; Sakurai, 1981; Amari et
al., 1999, 2006; Wheatland, 2006; (3) MHD relaxation method:
Chodura and Schlueter, 1981; Yang et al., 1986; Mikic and
McClymont, 1994; Roumeliotis, 1996; Valori et al., 2005, 2007;
Jiang et al., 2011, Jiang and Feng, 2012 (4) Optimization
approach: Wheatland et al., 2000; Wiegelmann, 2004, 2007; Inhester
and Wiegelmann, 2006; Wiegelmann and Neukirch, 2006 (5) Boundary
integral equation method: Yan and Sakurai, 1997, 2000; Yan and Li, 2006;
He and Wang, 2008; He et al., 2011.

As a stand alone method, the Boundary Integral Equation (BIE) method is one that allows us to evaluate the NLFFF
field at arbitrary points within the domain from the boundary data,
without the requirement to solve the field in the entire domain.
Moreover, {because the BIE model takes into account the asymptotic condition
consistently,} it allows us to only use bottom boundary data as the
boundary condition. This satisfies the current observational
condition and avoids assuming arbitrarily-prescribed lateral and top boundary data. The
BIE method for NLFFF was first proposed by Yan and Sakurai (1997,
2000), and many applications of BIE to solar events have been
implemented (e.g., Yan and Sakurai, 1997; Yan et al., 2001; Liu et
al., 2002; Yan, 2003). Later a new direct boundary integral
Equation (DBIE) method was proposed as an improvement to the BIE method. An optimization technique
has been applied to approximate the non-linear force-free field at any position
numerically. Compared with BIE, the complicated volume integration in Equations (17)
and (19) in the paper of Yan and Li (2006) were avoided. A series of test
cases and practical applications (Yan and Li, 2006; Liu et al., 2011, 2012; He and Wang 2008; He et al., 2011)
have been carried out to demonstrate the reliability and feasibility of DBIE.

Recently, with the launch of Solar Dynamic Observatory (SDO), {the Helioseismic and Magnetic
Imager (HMI; Schou et al., 2012) can provide the vector magnetogram which can be used as the high quality boundary data for coronal magnetic field reconstructions.  The Atmospheric Imaging Assembly (AIA; Lemen et al., 2011) can provide high resolution coronal structure images for the evaluation of any modeling techniques.}
Therefore, it is necessary to apply the DBIE method to real data by using high
resolution boundary data as validation. The previous BIE method was thought to be slow when carried out on entire 3D
domain (Schrijver et al., 2006; Wiegelmann, 2008) as the parallel algorithm was not implemented
though the BIE technique itself should be suitable for parallel computation.
In order to solve this problem, we implemented a Graphics Processing Unit (GPU) technique into our program
to accelerate the computing processes. The results show that this method is effective and suitable.

The NOAA 11158 was the first active region that produced X-class event in the current 24th solar cycle.  An X2.2 flare
event occurred  on 2011 February 15 at 01:44 UT.  Many studies have been carried out on this event, such as work
on the evolution of the magnetic field (Sun et al., 2012), research focusing on solar features (Schrijver et al., 2011),
extrapolations on the HMI, vector magnetogram (Wiegelmann et al., 2012),
evolution of relative magnetic helicity and current helicity (Jing
et al., 2012), non-potentiality of active region (Song et al.,
2013), and the work on the rotating sunspots of this region
(Vemareddy et al., 2012). Although most of these studies have the aid
of extrapolation methods, none of them have demonstrated the 3-D view of the
reconstructed coronal magnetic fields for this active region. The twin STEREO/A(head) and B(ehind)  spacecraft
(Kaiser et al., 2008) observe the Sun from multi-views, which provides us with a good
opportunity for a comprehensive comparison so that the physical process in the corona can be understood
correctly. {It should be mentioned that Su and van Ballegooijen (2012) compared a
NLFFF model with bright EUV features on the two sides of a solar polar crown prominence that erupted on
December 6, 2010 observed by STEREO B and AIA, since the channel was on the backside of the
Sun in STEREO A observations. DeRosa et al. (2009) compared other NLFFF models with observations including
STEREO A/B data for AR 10953 on April 30, 2007 but no comparison with STEREO images was shown. Sandman and
Aschwanden (2011) proposed a forward-fitting method with the stereoscopically reconstructed STEREO loops as known conditions. }

In this work, we apply DBIE method to active region NOAA 11158 on February 14 with the {HMI} vector magnetogram taken at 20:12 UT as boundary condition in order to understand the three-dimensional magnetic
configuration before the X2.2 flare event. We will present our reconstructed topology configuration of magnetic fields in the whole research region and electric current distribution in the central region. We then compare them with observations from both front view (SDO/AIA) and side views (STEREO A/B).

This paper is arranged as follows. Section 2 briefly introduces the DBIE method and GPU technique.
Section 3 shows the observations and Section 4 presents the reconstructed results.
Finally in Section 5 we draw our conclusions.

\section{Methods} 
      \label{method}

\subsection{Principle of DBIE} 
  \label{principle}

As an improvement of the BIE method, DBIE method also need to
satisfy the force-free field and divergence-free conditions (Yan
and Li, 2006):

\begin{eqnarray}
 & \nabla \times \textit{\textbf{B}}=\alpha\textit{\textbf{B}} \label{forcef}
 \\
 & \nabla\cdot\textit{\textbf{B}}=0 \label{divf}
\end{eqnarray}
with the boundary condition at $z=0$ magnetogram region (outside this magentogeam region a vanishing field is assumed):
\begin{equation}
\textit{\textbf{B}}=\textbf{B}_0 \label{bnd}
\end{equation}

At infinity, an asymptotic constraint should be employed to ensure
a finite energy content in the semispace above the Sun,

\begin{equation}
\textit{\textbf{B}}=\textit{\textbf{O}}(R^{-2}) \ \rm{when} \ R \longrightarrow \infty
\end{equation}
where $R$ is the radial distance. A reference function $Y$ is introduced in
this method
\begin{equation}
Y=\frac{\cos(\lambda\rho)}{4\pi\rho}-\frac{\cos(\lambda{\rho}')}{4\pi{\rho}'}\label{Y}
\end{equation}
where $\lambda$ is a pseudo-force-free factor depending on the
location of point $i$ only.
$\rho={[(x-x_i)^2+(y-y_i)^2+(z-z_i)^2]}^{(1/2)}$ is a distance
between a variable point  ($x, y, z$) and a fixed point $(x_i, y_i, z_i)$, ${\rho}'={[(x-x_i)^2+(y-y_i)^2+(z+z_i)^2]}^{(1/2)}$.

Combining with the force-free, divergence-free, boundary, and asymptotic
conditions, we obtain a direct boundary integral formulation (Yan
and Li, 2006)
\begin{equation}\label{DBIE}
\textit{B}_p(x_i,y_i,z_i)=\int_{\Gamma}\frac{z_i[{\lambda_{pi}}r\sin({\lambda_{pi}}r)+\cos({\lambda_{pi}}r)]\textit{B}_{p0}(x,y,0)}{2\pi[(x-x_i)^2+(y-y_i)^2+z_i^2]^{3/2}}dxdy
\end{equation}
where $r=[(x-x_i)^2+(y-y_i)^2+{z_i}^2]^{1/2}$, p=x, y, or z.
$\textit{{B}}_{p0}$ is the magnetic field on the photospheric
surface, and $\lambda_{pi}=\lambda_p(x_i,y_i,z_i)$ in place of $\lambda$ in Equation~(\ref{Y}), it is in principle
governed implicitly by the following expression:

\begin{equation}\label{lamd}
{\lambda_{pi}}^2=\frac{\int_\Omega Y(x, y, z; x_i, y_i, z_i, \lambda_{pi})[\alpha^2B_p+(\nabla\alpha\times \textit{\textbf{B}})_p]dxdydz}{\int_\Omega Y(x, y, z; x_i, y_i, z_i, \lambda_{pi})B_pdxdydz}
\end{equation}

Here $\lambda$ (denotes those $\lambda_{pi}$'s for short) has the same dimension as the force-free
factor $\alpha$. it is called pseudo-force-free factor. From Equation~(\ref{DBIE}), we can obtain
the magnetic field $\textit{\textbf{B}}$ if $\lambda$ is known. 
{A previous study on the property of $\lambda$ distribution by substituting the Low \& Lou (1990) solution
into the rigorous expression similar to Equation~(\ref{lamd}) was done for the BIE method (Li et al. 2004).
It was found that the $\lambda$ values that satisfy the condition at
some given point are not unique. However, this non-uniqueness of the $\lambda$ solutions does not influence
the computation of the field at that location, as demonstrated by numerical results. Obviously,
it is not practical to determine $\lambda$ from such an implicit expression~(\ref{lamd}). Yan and Li (2006) suggested to
make use of the Downhill Simplex method (Nelder and Mead, 1965) to find the suitable
$\lambda$ from a nonlinear programming problem. In this way the $\lambda$ is not obtained
from Equation~(\ref{lamd}) exactly but instead we look for a numerical solution
of the magnetic field. This is calculated from the given boundary
condition (3) together with the assumed asymptotic condition (4) and} satisfies the original force-free  (1) and divergence-free  (2) conditions approximately. Here the two stopping criteria of the procedure for the approximation of conditions (1) and (2) are  as follows.

\begin{equation}\label{fi}
f_i(\lambda_{xi},\lambda_{yi},\lambda_{zi})=\frac{|\textit{\textbf{J}}\times\textit{\textbf{B}}|}{|\textit{\textbf{J}}||\textit{\textbf{B}}|},\quad
\rm{with} \quad\textit{\textbf{J}}=\nabla\times\textit{\textbf{B}}
\end{equation}

\begin{equation}\label{gi}
g_i(\lambda_{xi},\lambda_{yi},\lambda_{zi})=\frac{|\delta\textit{\textbf{B}}_i|}{|\textit{\textbf{B}}_i|}=\frac{|\nabla\cdot\textit{\textbf{B}}|\Delta
V_i}{|\textit{\textbf{B}}|\Delta\sigma_i},
\end{equation}
and
\begin{equation}
f_i({\lambda}{^*_{xi}},\lambda{^*_{yi}},\lambda{^*_{zi}})=\min\{f_i(\lambda_{xi},\lambda_{yi},\lambda_{zi})\}
\end{equation}
\begin{equation}
g_i({\lambda}{^*_{xi}},\lambda{^*_{yi}},\lambda{^*_{zi}})=\min\{g_i(\lambda_{xi},\lambda_{yi},\lambda_{zi})\}
\end{equation}

We set the constraints like this:
\begin{equation}\label{figi}
f_i({\lambda}{^*_{xi}},\lambda{^*_{yi}},\lambda{^*_{zi}})\leq\epsilon_f,\quad g_i({\lambda}{^*_{xi}},\lambda{^*_{yi}},\lambda{^*_{zi}})\ \leq\epsilon_g,
\end{equation}
where $\epsilon_f$ and $\epsilon_g$ are sufficiently small
thresholds. Basically, $f_i(\lambda_{xi},\lambda_{yi},\lambda_{zi})$ is the sine of
the angle between $\textit{\textbf{B}}$ and $\textit{\textbf{J}}$,
which is used to evaluate the force-freeness. Similarly,
$g_i(\lambda_{xi},\lambda_{yi},\lambda_{zi})$ stands for the divergence of
$\textit{\textbf{B}}$.

Since we just gave a simple description about the approximation of $\lambda$ in the
previous work (Yan and Li 2006), and this may make some misunderstandings about our
method. Here we will state it in detail. As stated above, in the numerical procedure (Yan and Li, 2006),
we only need to control the force-freeness and divergence-free of the magnetic
field through Equation~(\ref{fi})~and~(\ref{gi}) approaching a minimum. The DBIE numerical procedure is
possible if the function $f_i$ can be calculated analytically. {In order to evaluate the right
hand side of (\ref{fi}) and (\ref{gi}), we need to know the space derivative of $\textit{\textbf{B}}$ from (\ref{DBIE}) and hence of $\lambda$.
This derivative is approximated by a first order finite difference. We evaluate $\lambda$ in the $\delta$-neighbourhood
of the point $r_i=(x_i, y_i, z_i)$, where $\delta$ is a very small positive fraction (typically one thousandth of the
pixel size). At an arbitrary} point in this small neighborhood it can be expressed as Equation~(\ref{lamdar}),
\begin{equation}\label{lamdar}
\lambda_p(r)=\lambda_p(r_i)+{\lambda_p}'(\xi)(r-r_i)
\end{equation}
which satisfies the Lagrange mean value theorem and $r_i<\xi<r$.
Since $|\delta|\ll1$ and $|r-r_i|\le\delta$, the zeroth-order
approximation is adopted. In our difference domain, we obtain
$\lambda_p(r) \approx {\lambda_p}(r_i)$. Here ${\lambda_p}(r_i)$
is a value of $\lambda_p(r)$ in the center of the small domain.
Then, any value of the function $\lambda_p(r_i)$ in the
infinitesimal neighborhood is known. The field $\textit{\textbf{B}}$ and the current $\nabla\times\textit{\textbf{B}}$ can then be evaluated over the point $i$. This is a practical and
rigorous numerical procedure.

However, Rudenko and Myshyakov (2009) wrote that they "think that this method for solving the extrapolation problem is incorrect" because they found that Yan and Li (2006) "unreasonably
drop these space derivatives" of ${\lambda}$ functions and "the resulting magnetic field will not be free-force".

Obviously, the comments in Rudenko and Myshyakov (2009) are incorrect as they have confused the DBIE representation of the
force-free field solution and the numerical approximation of the force-free field. It should be pointed out that the derivation of the DBIE is mathematically valid and rigorous. The problem is to find out how to obtain a numerical solution with the help
of DBIE.

As explained above, the strategy is not to solve Equations~(\ref{DBIE}) and~(\ref{lamd}) exactly but to find a numerical solution that satisfies the constraints (\ref{figi}) {and the boundary and asymptotic conditions (3-4)}. Alterenatively the original force-free and divergence-free equations (1-2) {together with boundary and asymptotic conditions (3-4)} is solved approximately. Therefore if one can construct numerically, the magnetic field distributions pointwise at any position that satisfies the constraints (\ref{figi}) {with the boundary and asymptotic conditions (3-4)}, one has already obtained a set of numerical solutions that are force-free and
divergence-free {with the given boundary conditions} approximately.

In the present work, our calculated results will further demonstrate the feasibility and validity of DBIE.

At the same time, the current density can be obtained pointwise:
\begin{equation}\label{J}
\textit{\textbf{J}}=\nabla\times\textit{\textbf{B}}
\end{equation}

As one of the advantages of DBIE, it is a pointwise method, which
can calculate the magnetic field and current density in any point
above the photospherical boundary from the procedure. {However, it should be noted that a vector magnetogram with all three field components is more than a force-free field needs to be uniquely determined. The present DBIE employs the vector field in the reconstruction. Therefore the boundary data should  satisfy compatibility relations in order to be consistent with a force-free corona. The inconsistency and errors contained
in the vector magnetogram data will cause errors in the reconstructed field. The ignorance of the boundary field
${\bf B}_{0}$ outside of the magnetogram area would also have influence to the reconstructed field. In practice, the
truncation of the magnetogram data should be chosen to approach zero as ${\bf B}_{0}$ vanishes outside of the
magnetogram area. Nevertheless, the net flux of ${\bf B}_{0}$ in Equation (3) over the boundary magnetogram area does
not need to be zero as shown in the derivation of the BIE (Yan and Sakurai 2000) or DBIE (Yan and Li 2006). }

\subsection{GPU Technique} 
  \label{GPUtech}

With more and more advanced telescopes launched into
space, higher quality images have become available.  On one hand, high
resolution images provide more clarity to the detail of the Sun and this will help
us to study the nature of the Sun in more detail. On the other hand, the vector magnetogram used
as the boundary condition of NLFFF methods is getting larger
which will vastly increase the amount of computation for each method. For the BIE method, it is necessary to solve the
computing speed problem and apply it to real data by using high
resolution boundary data. The BIE or DBIE method are in principle suitable for
high-performance parallel computing. However, in the previous
work, the implementation of BIE with parallel computing on high
performance computers was not carried out. Therefore BIE would
be slow when extrapolating NLFFF from boundary data with computing
grids compatible with current observations. DBIE is expected to make an improvement
(Schrijver et al., 2006; Wiegelmann, 2008). Hence we will adopt a suitable parallel computing technique for DBIE.

In recent years, the graphic processing technique
has become prevalent in general purpose calculation. We utilized Graphics
Processing Units (GPUs) in our program. The results turn out to be
effective and suitable.  We can replace a CPU cluster
consisting of tens of CPUs with just one GPU board fixed on a personal
computer. This convenience has
profound meaning on the promotion of the application for the DBIE method.

A GPU is composed of high-performance multi-core processors capable
of very high computation and data throughput (Zhang et al., 2009).
The GPU's powerful parallel computing ability to process the
integration operation can be applied to the DBIE method. Parallel
computation of the DBIE-NLFFF extrapolation algorithm is performed
through GPU with shared memory accessing optimization under Linux
system and a Compute Unified Device Architecture (CUDA) compiler.
The calculation is operated through an Intel CPU @ 3.40 GHZ and
NVIDIA Geforce GTX 480 graphics device with NVIDIA CUDA 4.2 on a
personal computer.

The platform employed in this work is a 4 core CPU and one GPU
machine. The main part of the program is the integral operation in
Equation~(\ref{DBIE}). The iteration part is executed mostly in
the CPU cores and the data is transferred between CPU and GPU. In
order to reduce the latency and improve the occupancy of the
procedure, we need to reduce the data exchange in the global
memory between CPU and GPU, and allocate reasonably the numbers of
the $\textit{Thread}$ and $\textit{Block}$. The number
is not fixed and there are some allocation rules which may improve
the speed. Generally speaking, there are some tricks to allocate.
The thread number is a multiple of 32, which can improve the
memory coalescing of the procedure. For different sizes of data,
the number is different, the larger the better, since it can
improve the occupancy. We can adjust the thread between 128 and
256, and then change the block number gradually. Meanwhile, we
should make sure the block number is larger than the
multi-processor, which can guarantee no multi-processor is empty.
In our work, the numbers of $\textit{thread}$ and $\textit{block}$
are 128 and 80, respectively, which provide good
allocation in our procedure. In addition, we utilize the shared memory for optimizing our program to
improve the computational speed. This can reduce
volume of the output data transmission from GPU to CPU.

According to Equation~(\ref{DBIE}), in the numerical  procedure the
magnetic field of an arbitrary point $i$ in the semispace above the
boundary can be expressed as follows (Yan and Sakurai, 2000).
\begin{equation}\label{DBIE-dis}
\textit{\textbf{B}}_i=\sum_{e=1}^{Ne}\sum_{j=1}^{9}\left [\int_{-1}^{+1}\int_{-1}^{+1}YN_k(\xi,\eta)J(\xi,\eta)d\xi d\eta \right]  \textit{\textbf{B}}_{j}^{e}
\end{equation}
where the boundary has been subdivided into $N_e$ 9-nodes elements with boundary data known over each node, $N_k(\xi,\eta)$ is the shape function, $J(\xi,\eta)$ denotes the Jacobian, and
$\textit{\textbf{B}}_{j}^{e}$ indicates known nodal field values as provided by the boundary condition
similar to $\textit{B}_{p0}$ in Equation~(\ref{DBIE}).

\begin{figure}
\centering
\includegraphics[width=4.8in]{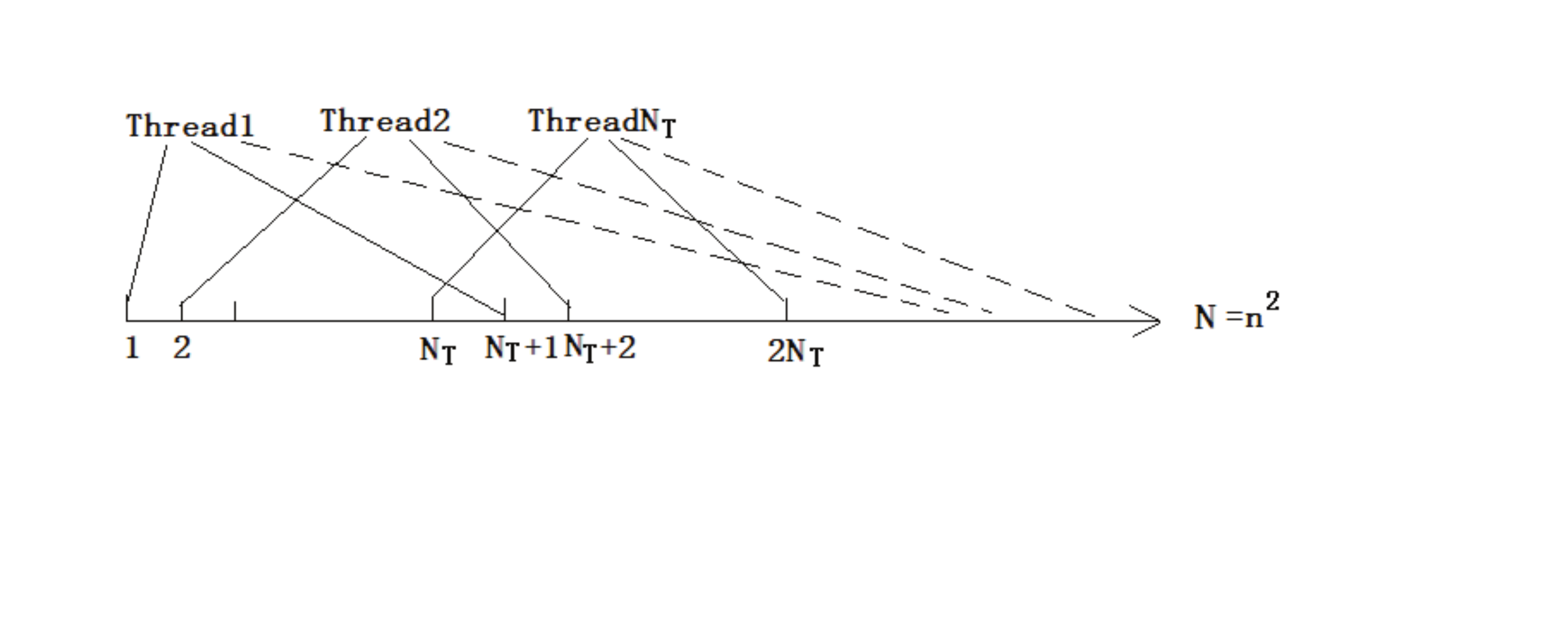}
\caption{The stretch of the GPU assignment allocation. The data
scale is presented by N that equals to the square of boundary
grids n. $N_T$ indicates the number of GPU thread. The lines
present the parts of assignment that are put into corresponding
GPU threads.}\label{gpu}
\end{figure}

For clarity, we simplify this equation as Equation~(\ref{DBIE-dis2}), where $N$ equals
to $n\times n$ namely the number of grid nodes of the boundary
condition. We allocate our GPU assignment like Figure~(\ref{gpu}),
the number of $thread$ is expressed by $N_T$, the boundary grids
are marked as $1, 2, ..., N_T, N_{T+1}, N_{T+2}, ..., 2N_T,...,
N$. The boundary data are put into each $thread$, and $threads$
are put into $blocks$. Thus, our data parallelization is realized.
Then we get the summation of the data in each $thread$, then add
the summation results in each $block$.

\begin{equation}\label{DBIE-dis2}
\textit{\textbf{B}}_i=\sum_{k=1}^{N} a_{ik} \textit{\textbf{B}}_{k}
\end{equation}

A series of numerical tests indicate that the GPU-accelerated DBIE program is almost 1000 times faster than
the original DBIE, which is including the hardware update, difference
of the compiler, instruction optimization and GPU's effect. The
total computation cost can be expressed as $\textit{O}(n^2m^3)$
(Yan and Li, 2006), {which has to be multiplied with the
number of iterations to minimize $f_i$ and $g_i$ in Equation (12),} where $n\times n$ is the boundary nodes and
$m\times m\times m$ expresses as the cubic grids. As Figure~\ref{gpu} shows us that point $i$ in the semispace above the boundary needs to do the integral operation to the $n\times n$ boundary grids
$\textit{B}_{p0}$ (in Equation~(\ref{DBIE})). We only apply the GPU
acceleration into making this $n^2$ part parallelized.
However, the internal grids (or $m^3$ part) parallelization, namely the
number of points $r_i$ are not involved yet. So further acceleration
could combine CUDA with other parallel computing techniques such as Message Passing
Interface (MPI) to realize Muti-GPU parallelization.

Before we apply the present DBIE method to analyze the practical problems we first compare
it with a semi-analytical solution for NLFFF. As no iteration was performed to determine
the set of factors by the BIE method in the comparison against the analytical force-free-field
models of Low and Lou (1990). It was expected that completion of the iteration by DBIE will
greatly reduce the computation time and should be feasible (Schrijver et al., 2006). Here we
just adopt the Case \uppercase\expandafter{\romannumeral2} in Schrijver et al. (2006),  i.e.,
only bottom boundary data are used because this type of the boundary condition is close
to the case of the Sun. The boundary size and the five evaluation metrics $C_{vec}$, $C_{cs}$,
${E^{'}}_{n}$, ${E^{'}}_{m}$ and $\epsilon$ are the same as in Schrijver et al. (2006). The results and
the comparison with other methods are shown in Table~\ref{tb}. It can be seen that after iteration
by the present DBIE method, the metrics have been significantly improved as compared with the boundary
integral method without an iteration and similar results have been obtained as compared with
other methods.
\begin{table}
\caption{ Evaluation of metrics for the present DBIE and other
methods.} \label{tb}
\begin{tabular}{lccccc}     
  \hline                   
Only lower boundary provided, entire volume\tabnote{The parameters are the same as in Case \uppercase\expandafter{\romannumeral2} in Schrijver et al. (2006) with Low \& Lou (1990) solution: n=3, m=1, l=0.3, $\Phi=4\pi/5$ on a 192 $\times$ 192 pixel grid centered on the
$64\times64\times64$-pixel test region.} & $C_{vec}$ & $C_{cs}$  & ${E^{'}}_{n}$ & ${E^{'}}_{m}$ & $\epsilon$\\
  \hline
Exact solution (Low \& Lou, 1990)&1& 1& 1& 1& 1\\
\hline
Weighted Optimization Method (Wiegelmann)\tabnote {Data from Table~I of Schrijver et al. (2006).} & 1.00 & 0.57  & 0.86 & -0.25 & 1.04 \\
Optimization Method (McTiernan)$^2$ & 1.00 & 0.51  & 0.84 & -0.38 & 1.04 \\
Magnetofrictional Method (Valori)$^2$ & 0.99 & 0.55  & 0.75 & -0.15 & 1.02 \\
Grad - Rubin - like Method (Wheatland)$^2$ & 0.99 & 0.58  & 0.69 & 0.13 & 0.96 \\
Grad - Rubin - like Method (R$\acute{e}$gnier)$^2$ & 0.94 & 0.28  & 0.49 & -1.7 & 0.74\\
Boundary Integral Method (no iteration)$^2$ & 0.97 & 0.41  & -0.02 & -14. & 1.00 \\
  \hline
Upward-layered DBIE Method (He)\tabnote{Data from Table~4 of  He \& Wang (2008).} &0.97&0.65&0.077&12.4&1.06\\
Present DBIE Method & 0.99 & 0.52 & 0.83 & -0.53 & 1.08\\
  \hline
\end{tabular}
\end{table}

\section{Observations} 
      \label{obs}

The NOAA 11158 was the first active region that produced an X-class event in the current 24th solar cycle.
There were many C-class and M-class flares in this active region during its passage over the solar disk in February 2011.
The largest, the X2.2 flare event occurred on February 15 at 01:44 UT. Several studies have been carried out on NOAA
11158 (Schrijver et al., 2011; Sun et al., 2012; Wiegelmann et al., 2012). The proposed GPU-accelerated DBIE is
applied to reconstruct the coronal magnetic field from the vector magnetogram
taken on 2011 February 14 at 20:12 UT from SDO/HMI. This is combined with observations from the SDO/AIA and
the two STEREO/Extreme Ultraviolet Imager (EUVI) instruments (Howard et al., 2008; W$\ddot{u}$lser et al., 2004) to
present a stereoscopic investigation of the coronal magnetic fields in order to understand the X2.2 flare event.
We {average} the boundary data from 360 ${\rm km~pix^{-1}}$ ($0.5''$) to 720 $\rm km~pix^{-1}$ (about $1''$), which has $300\times300$ grid points to be used as the boundary condition.
In order to make a comparison with previous work, we also pay attention to the
central $250\times200$-pixel area covering the main features of the active region, with the vertical grid
spacing matching the horizontal spacing.

\begin{figure}    
   \centerline{\includegraphics[width=0.8\textwidth,clip=]{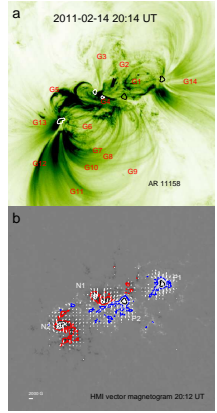}
              }
\caption{ The observations from SDO. (a) is the EUV image in 171
$\AA$ of NOAA AR 11158 from SDO/AIA on 2011 February 14 at 20:14
UT. The EUV loops are divided into 14 groups and
marked from G1 to G14. The field of view is about
$300''\times300''$. (b) is the vector magnetogram from SDO/HMI at
20:12 UT. The horizontal fields are presented by using arrows with a length scale of 2000 G shown by the white bar. The
vertical fields are plotted by contour map at $\pm1000, 2000$ G.
P1, N1, and P2, N2 present two pairs of reversed polarities. Red
indicates negative and blue is positive.
                      }
   \label{magmap}
   \end{figure}

Here, we used the HMI vector magnetogram as the boundary data with three components of the magnetic field
shown in Figure~\ref{magmap}(b), and the two main pairs of bi-polarity are marked as P1, N1
and P2, N2 there. The cutout data used for the force-free field
modeling has been mapped to a local Cartesian coordinate.

Considering the precision of our method largely depends on the
boundary condition, solar magnetic field measurement suffers
from several uncertainties (McClymont et al., 1997). Wang et al. (2001)
proposed a method to remove the $180^\circ$ ambiguity and
make the boundary data reduction for the BIE method. Here we apply
this data reduction method to the boundary data in the present study.

\begin{figure}    
\centerline{\includegraphics[width=1.0\textwidth,clip=]{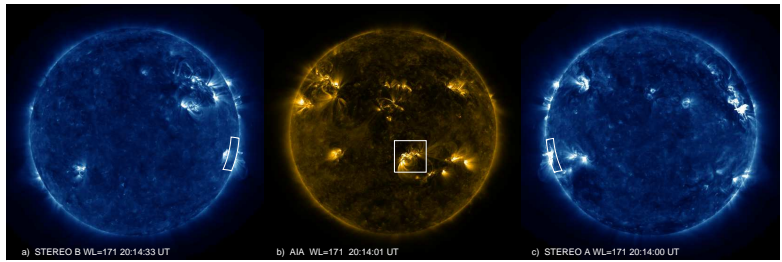}}
\caption{ Full disk maps from AIA (b), STEREO A (c), and STEREO B
(a) in 171 \AA. The research region for extrapolation is marked
by a white square in (b), and corresponding domain is marked in
(a) and (c).} \label{fulldisk}
\end{figure}

After reconstructing the coronal magnetic field using the GPU accelerated DBIE method,
we compare the modeling results with the EUV images of AIA and STEREO/EUVI from
three points of view in order to quantify to what extent they
correctly reproduce the coronal magnetic field configuration. To
co-align the cutout vector magnetogram with AIA images we carried out
a correlation analysis between $B_z$, from the
original vector magnetogram, and $B_{los}$ from
the full disk LOS magnetogram. Then the location of the rectangle research
region (shown as the white squares in Figure~\ref{fulldisk}) is
determined in the full disk SDO/HMI magnetogram. According to the
SDO data analysis guide, we align HMI data with AIA data
and obtain the cutout AIA image (shown as Figure~\ref{magmap}(a)) and
the location of our research region. In order to determine the
location of the research region in STEREO images, the coordinate
conversions between Stonyhurst heliographic and
heliocentric-cartesian are adopted (Thompson, 2006). Thus the
reconstructed results viewed from three different points of view
are shown aligned with the EUV background with accurate locations.

\begin{figure}    
\centerline{\includegraphics[width=0.8\textwidth,clip=]{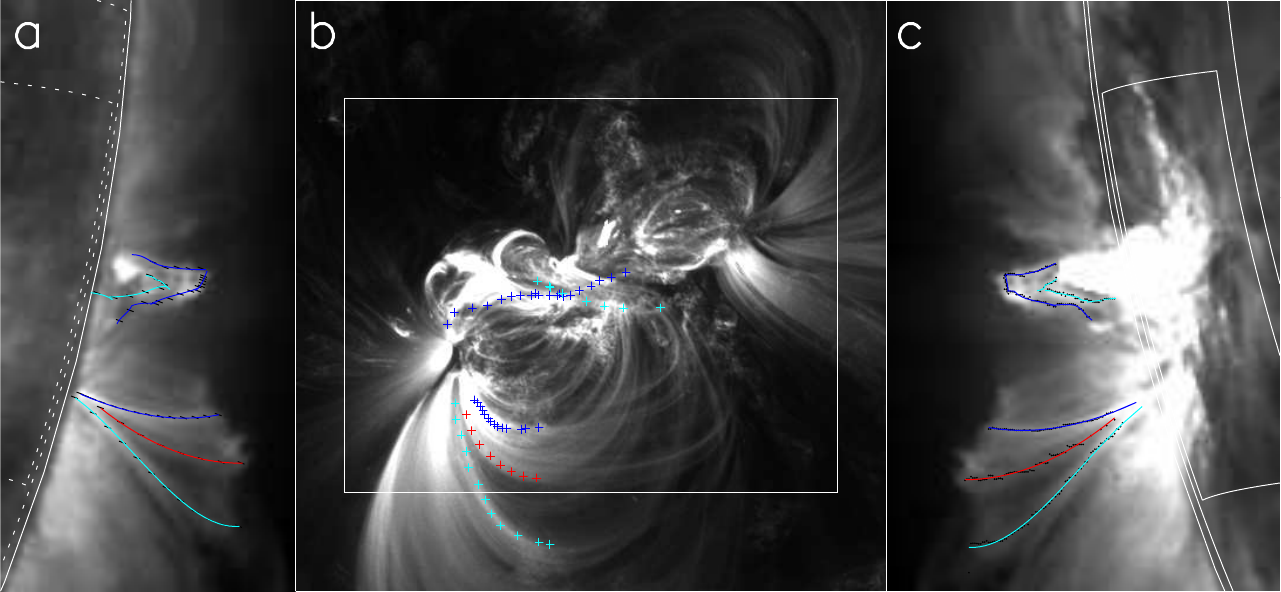}}
\caption{ Corresponding positions of the stereoscopically reconstructed coronal loops from
three different points of view. It presents the loop features in lower
part (blue, red, and light blue cross points) and middle part (blue and
light blue cross points) of the AIA image (b), with the corresponding loop features denoted by the
same color lines in the STEREO A (a) and B (c) images.}
\label{EUVI3}
\end{figure}

Before we compare our reconstruction results with observed EUV loops, we need to determine the same
features in AIA image and the two STEREO/EUVI images.
For a coronal loop in the STEREO A image shown as the light blue line in the bottom of Figure~\ref{EUVI3}(c),
we apply a Gaussian-fitting to the cross section of the loop and find the brightest point along this cross section.
Then we select a number of cross sections along this loop and connected these points together.
Thus, we get the $'skeleton'$ of the loop. Through the coordinate conversions (Thompson, 2006),
some selected points along the $'skeleton'$ line are projected to the image of STEREO B in
Figure~\ref{EUVI3}(a) and these projections seen as {short black
bars} (whose length seen in the AIA image is nearly the same as the
length of the research region). By using the same method as STEREO A image, we get
the $'skeleton'$ line of the loop in the image of STEREO B and we obtain
the points of intersection. We convert these points of
intersection in STEREO B image and points in STEREO A image to AIA image in
Figure~\ref{EUVI3}(b). Thus, the points of intersection from STEREO A and
B lines are obtained and are marked as many cross-shaped points in the
AIA image. We can thus obtain the stereoscopically reconstructed coronal loops. We apply this method mainly in
some higher altitude structures which could be seen from both STEREO A/B EUVI instruments.
Therefore, the comparisons below will take into account these obvious
higher altitude structures.

\begin{figure}    
\centerline{\includegraphics[width=1.0\textwidth,clip=]{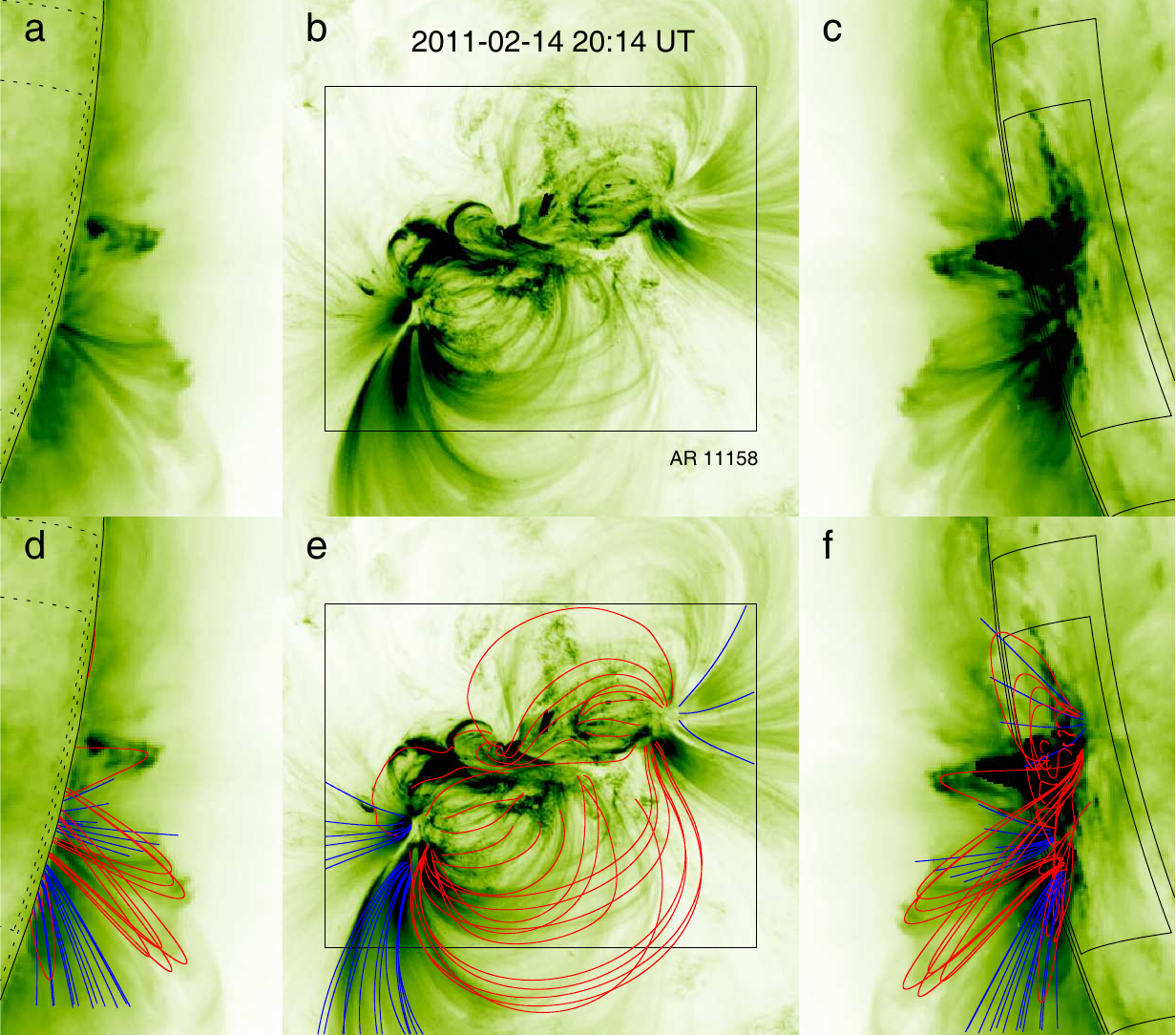}}
\caption{ Comparison between EUV images and reconstructed
results. The first row presents the EUV images in 171 \AA
from STEREO B (a), AIA (b), and STEREO A (c). The same images
superimposed with extrapolated magnetic field lines are presented
in the bottom row images. The red lines show the closed
extrapolated magnetic field lines, the blue lines are the open
magnetic field lines which extend to the outside of boundary area
that is $300''\times300''$. The black squares in (b) and (e)
present the $250''\times200''$ domain which contains the main
features of EUV structures. The outer square in (c) and (f) image
presents the boundary area and inner one is the same as the
squares on the AIA images. The same region in the backside of the
Sun in (a) and (d) are presented by dotted lines.} \label{mainmap}
\end{figure}

We present some selected EUV bright loop structures in 171 {\AA}, and divide the
EUV features into groups marked from G1 to G14 (see Figure~\ref{magmap}(a)).
G1, G2, and G3 are three groups corresponding to the EUV loops on the top of the research region and connect the
magnetic polarities P1 and N1. When we determine the EUV vertical structures rooted at the edge
of the solar disk and stretching out of the disk from the side views by the 3-D stereoscopic technique
we find they are out of the interested region. So loops G1 could not be seen {from the views at  the solar limb} and they are in lower altitudes. G4 is the kernel region where stretched loops along the
polarity inversion line (PIL)  are observed and the flare event occurred. So we present not only the reconstructed magnetic field lines but also the electric current lines. G5 and G6 present some lower small loops which can be distinguished just {from the view on the solar disk}.
G7, G8, and G9 are large loops connecting N2 with P2. G10 and G11 are
also the large loops connecting N2 but with P1. These bundles of loops could be seen from
three different points of view. G12, G13, and G14 are the open loops extending
to the outside of the interested region and rooted from N2 and P1 respectively.

\section{Reconstructed Results} 
      \label{result}

The extrapolation code is based on the GPU-accelerated DBIE
method. Alignments between the extrapolated field lines and EUV
images in 171 $\AA$ from the SDO/AIA, and twin STEREO/EUVI
instruments are presented (Figure~\ref{mainmap}). The DBIE method
is performed to reconstruct the 3D magnetic field structures of
the region NOAA 11158 in the corona. The red lines in
Figure~\ref{mainmap}(e) show the calculated closed magnetic field
lines. Blue lines present the calculated open magnetic field lines
which extend to the outside of our computational domain.
Figure~\ref{mainmap}(d) and (f) show us the counterparts
reconstructed field structures from STEREO B and A, respectively.

\begin{figure}    
\centerline{\includegraphics[width=0.9\textwidth,clip=]{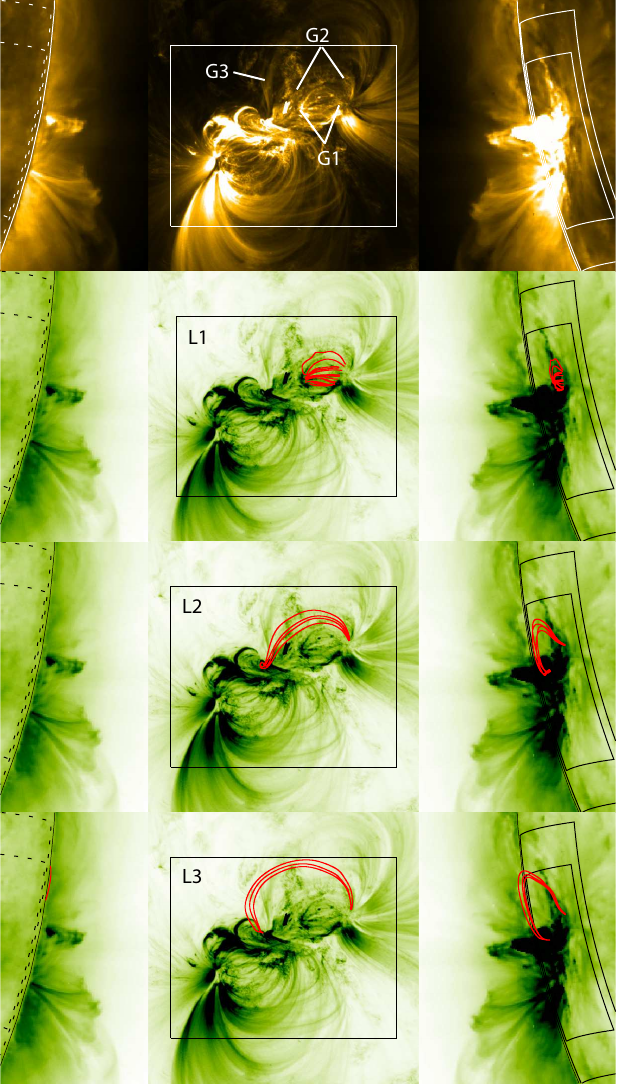}}
\caption{ The comparison between calculated magnetic field lines L1, L2, L3 and EUV loops in 171 $\AA$ G1, G2, G3.}
\label{G123}
\end{figure}

Figure~\ref{G123} presents the decomposed comparison between EUV
{loop groups G1, G2, G3 and our reconstructed magnetic field lines.
The L represents the group of reconstructed magnetic field lines, G represents the group of EUV loops. }
The first row of Figure~\ref{G123} presents the EUV patterns in 171 \AA.
G1 consists of a series of small lower loop structures, which are not seen from STEREO B and show really good
agreement with our reconstruction results
L1 {from the view on the solar disk}. This bundle of loops connects P1 to the
relatively weaker negative magnetic polarities between P1 and N1.
G2 is the same as G1. It can also be seen from STEREO
A. It is worth mentioning that the left footpoints of calculated
lines in L2 show a helical structure and this
agrees well with the EUV background G2 around the negative
polarity N1. According to the method stated in Section~\ref{obs},
the EUV loops of G3 are lower than the vertical structures
stretching out of the solar disk. Compared with calculated magnetic field lines
in L3, it seems to be consistent with this, namely has a lower
altitude seen from STEREO A and could not been seen from STEREO B.
It is the largest loop bundle connecting P1 with N1.

\begin{figure}    
\centerline{\includegraphics[width=0.9\textwidth,clip=]{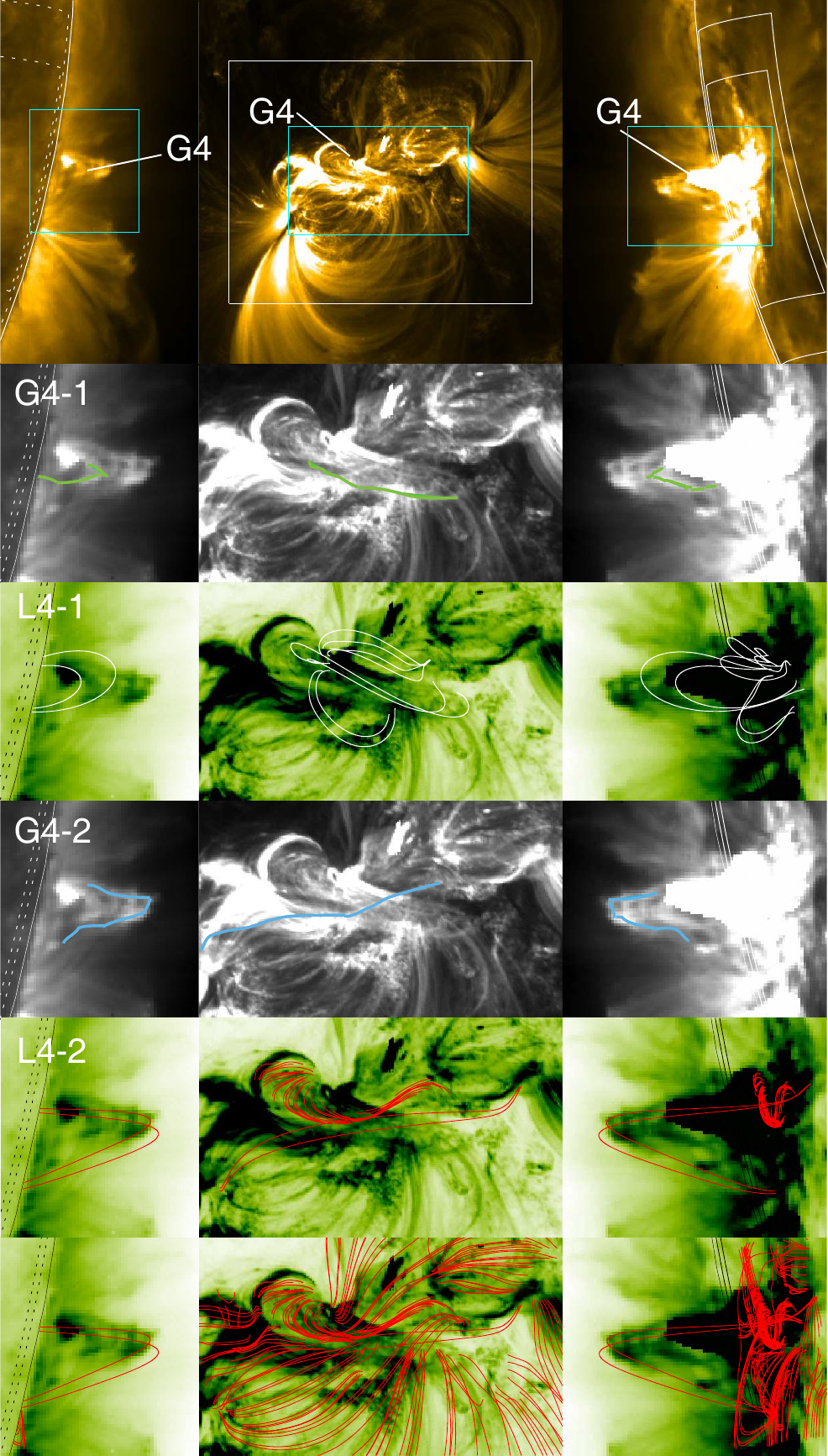}}
\caption{ The three blue squares in top panel represent the same
region, within which the results are displayed in the following panels,
around the polarity inversion line (PIL) from three different views for G4.
G4-1 and G4-2 are the close-up views of G4. The green lines
in G4-1 agrees with the calculated electric current lines (white)
in L4-1. The blue lines in G4-2 agree with the higher-lying calculated
magnetic field lines (red) in L4-2. The highly-twisted short lower-lying field lines in L4-2 form a S-shape co-spatial with  a filament channel there along the PIL. At bottom panel the calculated magnetic field lines are shown in detail to demonstrate
the pivot location of L4-2 to all other surrounding coronal structures L1, L2, L3, L5, L6, and the one-side footpoints of L7,
L8, ..., L12, and L14, etc. }
\label{G4}
\end{figure}

G4 is a group of EUV loops in the region around PIL and between magnetic polarities P2 and N1 (see Figure~\ref{G4}). There
is a strong magnetic shear and should contain a large amount of magnetic free energy
around the PIL, which is the most important region for understanding the physical processes of solar eruptions (see the
blue box in first row images of Figure~\ref{G4}). This region has relatively complex structures seen from AIA, and
there are also vertical structures stretching out of the edge of the solar disk (seen in the boxes in the STEREO side view images).
We determine the correspondence of the features in all three images. {Around the PIL, there
are some observed EUV loops connecting P2 with N1. Our extrapolation has obtained a series of small and low calculated
magnetic field lines along the PIL and connecting the regions on both sides, which are agreeable with the EUV
loop structures in general and the filament structure marked in Figure 1 in Sun et al. (2012), as shown in the last two
rows in Figure~\ref{G4}. However, we did not obtain higher-lying calculated magnetic field lines over the filament channel connecting footpoints with opposite magnetic polarities P2 and N1. Nevertheless, the current lines connecting P2 with N1 are
found and plotted in white marked as L4-1, which can be calculated from Equation~(\ref{J}). The location of the same
corresponding EUV loops in three view aspects are also marked in G4-1 shown as  green lines. In the STEREO A image, the upper
half descending part of this loop structure is blocked by the saturated patch in the 171 \AA ~detector, but the other visible part is in good agreement with the central electric current lines in L4-1. The highest structure above the one in G4-1 is shown as blue line in G4-2 representing large scale loops connecting P1 to N2 along the PIL in the center region of G4 loops. L4-2 shows
bundles of magnetic field lines in the kernel region. One bundle of field lines is located higher than the electric current lines
and they show very good agreement with the coronal loop denoted as the blue line in G4-2 in both front and side views. Other
bundles of lower-lying and short twisting field lines in L4-2 is connecting P2 and N1 along the PIL and co-spatial with the
S-shaped filament channel, where some EUV strands in the dark filament channel were also shown in Sun et al. (2012). The
highly-twisted short lower-lying field lines in L4-2 are in a pivot location to all other surrounding coronal
structures L1, L2, L3, L5, L6, and the one-side footpoints of L7, L8, ..., L12, and L14, etc. Therefore they must have played a
key role for the occurrence of the X2.2 flare event. In the STEREO A image, it can be seen that those lower-lying field lines in
L4-2 form a twisted arcade structure along the PIL where the filament is located. It should be noted that while the strand in
the left part of the S-shaped filament channel may be lower-lying and related to the filament, the right part EUV bright features
along the PIL, marked as a filament in Figures 1 and 2 in Sun et al. (2012), may not be necessarily all lower-lying, as it agrees
almost identical to the coronal loop in G4-2 that is clearly stereoscopically resolved as a high-lying coronal bright feature seen
in STEREO A/B EUV images.  At least one cannot simply attribute all those EUV bright features along the PIL in the filament
channel to manifestation of a filament although the filament could be located there. }

{Theoretically the field- and current lines should be identical for a force-free field but in practical situation there may exist 
discrepancy. For the current lines in L4-1 the averaged misalygnment angle between the fields and the current lines is 
$13.6^{\circ}$. This may be due to the inconsistency with force-freeness and errors
contained in the boundary data in the PIL region.  It should be pointed out that by the DBIE method the angle between the
current {\bf J} and the field {\bf B} is mostly less than $5^{\circ}$ with an average value of $4^{\circ}$ but there do exist
some points where the angles are large, whereas the relative flux error factor $g_i$ always has a maximum value of about
0.5\% when comparing with the exact solution (e.g., Fig.5 in Yan and Li 2006). In the present case for
$250\times200\times100$ internal grid points, the averaged values are respectively $<f_i>=0.078$ (or the averaged angle between {\bf B} and {\bf J} is less than 4.5$^{\circ}$) and $<g_i>=0.00067$. Physcially the coronal EUV loops are controlled
by the observed photospheric vector magnetogram data which are not necessarily force-free. Therefore there is no
guarantee that the observed coronal EUV loops are always consistent with a force-free field solution, especially for the solar flare
or coronal mass ejection events. Nevertheless the comparison between the calculated field lines and the observed coronal loops
would reveal the quality of the extrapolation.}

\begin{figure}    
\centerline{\includegraphics[width=0.9\textwidth,clip=]{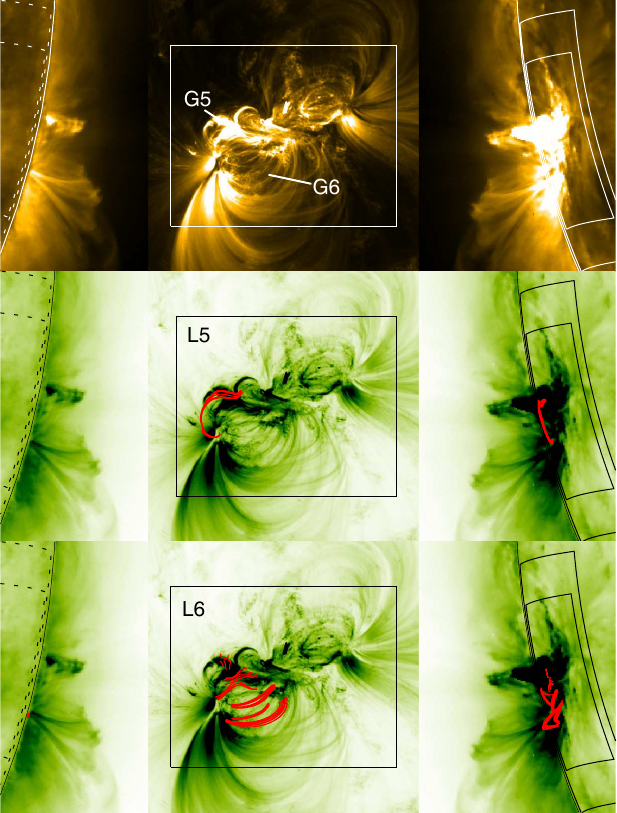}}
\caption{The comparison between calculated magnetic field loops
L5, L6 and EUV background patterns in 171 $\AA$ G5, G6.}
\label{G56}
\end{figure}

G5 and G6 are relatively low-lying EUV bright loops (see Figure
\ref{G56}). From the views of STEREO A, G5 shows some highlight
structures from which we cannot distinguish the details. The
corresponding calculated magnetic field lines in L5 are lower and cannot be seen
from STEREO B, which connect negative polarity N2 with the
positive polarity between N1 and N2. It is important to note that there were a series of drastic solar activities
before our selected extrapolated time in this region of G5. There
was a large CME on February 14 at 18:00 UT, with the associated flare
M2.2 at 17:20 UT from this site. G6 corresponds to a series of lower loops.
The calculated magnetic field lines L6 are also lower-lying and are qualitatively in good agreement
with observations from the AIA view of G6. From side views, they
mix with the background and no obvious features could be seen.

\begin{figure}    
\centerline{\includegraphics[width=0.8\textwidth,clip=]{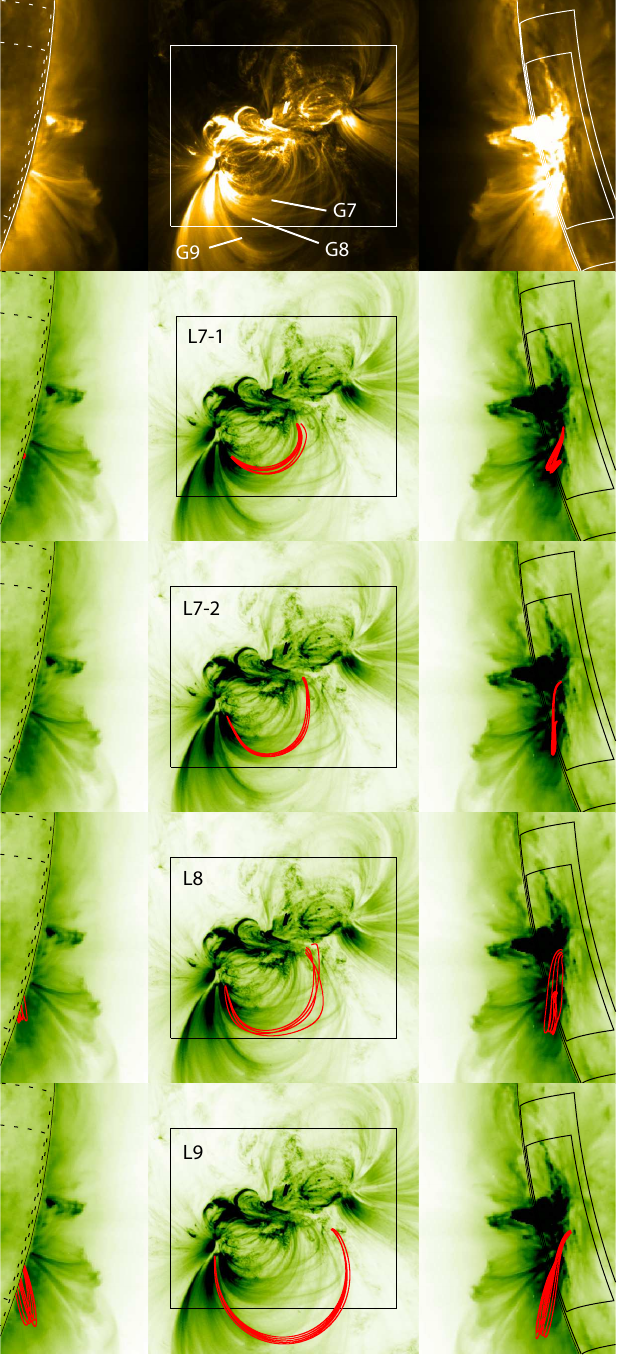}}
\caption{The comparison between calculated magnetic field loops
L7-1, L7-2, L8, L9 and EUV background patterns G7, G8, G9 observed
at 171 $\AA$.} \label{G789}
\end{figure}

G7, G8, and G9 consist of a series of coronal loops with different
length scale (see Figure~\ref{G789}). These loops all originate
from polarity N2 and their endings are around the polarity P2. The calculated L7, L8 and L9 lines
are all in good spatial agreement with the EUV G7, G8 and G9 loops {from the AIA view on
the solar disk}. The calculated field lines, L7-1 and L7-2 are lower and shorter, which
can be seen just in STEREO A not STEREO B. The calculated
lines are in good agreement with the EUV loops from the AIA image.
L8 connects N2 to P2 and has good spatial co-alignment with G8 in the AIA image.
However, their side projections still mix with the background and cannot be distinguished.
L8 also shows poor spatial co-alignment with the high altitude STEREO A/B EUV loops.
G9 is one of the largest loop bundles at the bottom of this region. As the loops grow, their
altitude increases. This could be seen from the side
view in the last row of Figure~\ref{G789}. L9 is in good agreement with G9 {from the front view} but show poor spatial co-alignment with the high altitude STEREO A/B EUV loops.

\begin{figure}    
\centerline{\includegraphics[width=0.9\textwidth,clip=]{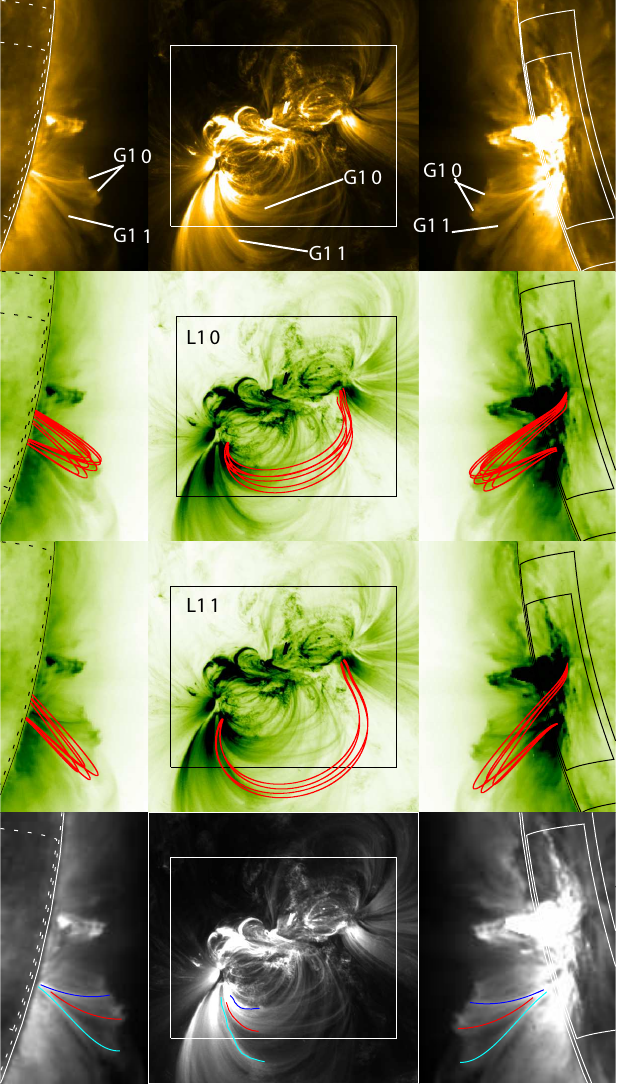}}
\caption{The comparison between calculated magnetic field loops
L10, L11 and EUV background patterns G10, G11 at 171 $\AA$. The
bottom row images give the corresponding positions of the coronal
loops G10 (blue and red) and G11 (light blue) from three different
points of view.} \label{G1011}
\end{figure}

We expect to find the coronal loops with not only the EUV patterns in AIA images but also in STEREO images.
Figure~\ref{G1011} shows a series of loops connecting N2 to P1.
Their corresponding structures are shown, from different angles, in
the first row of Figure~\ref{G1011}. The correspondence of these EUV
loops at different points of view are confirmed by our method
stated in Section~\ref{obs} and shown in the last row of
Figure~\ref{G1011} in different colored lines. However, we just
confirm the correspondence of half loops and the other half cannot be determined.
Nevertheless, this comparison validates our reconstructed
results. We could see that the calculated magnetic field lines L10
agree well with the EUV loops of the left parts of G10. It overlaps
with some parts of G7 and G8, but its ending is different. G11 also connects the spots
N2 with P1 and has the same situation with G10. It overlaps with some parts of G9. It can be seen that the
configurations of the calculated lines L10 and L11 are coincident
with the coronal loops G10 and G11 in both {front view from AIA and STEREO A/B side views.
Although the calculated field lines and observed EUV coronal loops agree with each other globally, they do not follow the same
trajectories. In order to make a quantitative comparison for the three stereoscopically reconstructed loops, we calculated the angles between the tangent vectors along reconstructed loops and the calculated fields there, and obtain the averaged misalignment angles of 16.6$^{\circ}$, 17.8$^{\circ}$, and 18.3$^{\circ}$ for three reconstructed loops in the middle
panel at last row in Figure~\ref{G1011} from bottom up. These values demonstrate the deviation from the force-freeness along these loops, which are quite good with a factor of about two smaller than those given by other NLFFF models yielding overall misalignment angles of $24^{\circ} - 44^{\circ}$ (DeRosa et al. 2009) and at the same order as a forward-fitting model using stereoscopically reconstructed loops as constraints (Sandman and Aschwanden 2011).}

\begin{figure}    
\centerline{\includegraphics[width=0.9\textwidth,clip=]{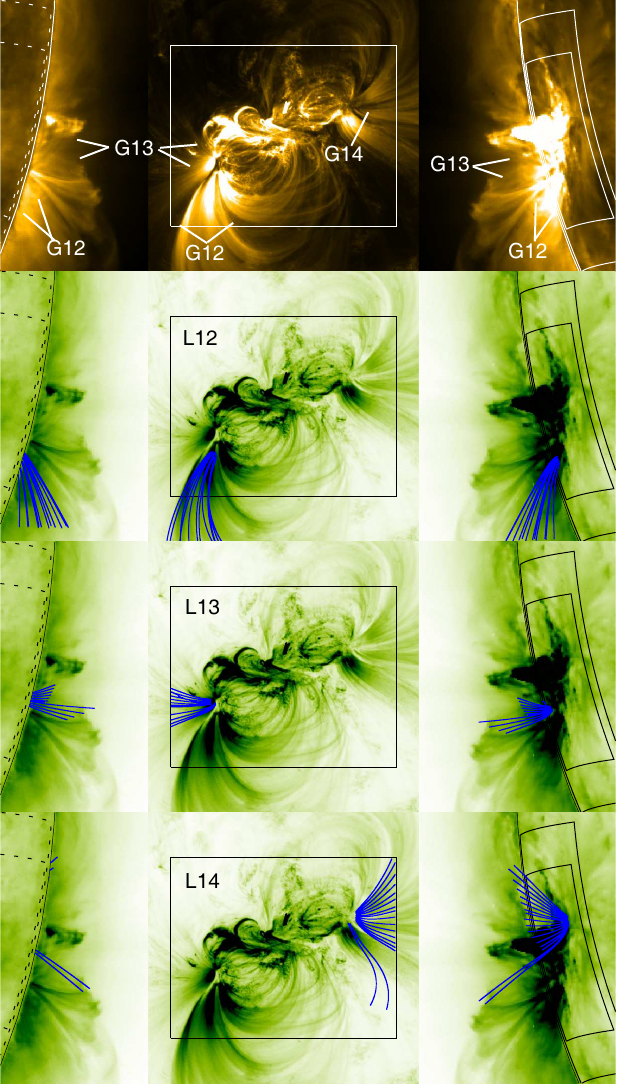}}
\caption{ The comparison between calculated magnetic field lines
that extend to the outside of boundary condition L12, L13, L14 and
EUV loop patterns G12, G13, G14 at 171 \AA.}
\label{G121314}
\end{figure}

G12, G13, and G14 consist of a series of large coronal loops which
are open to the outside of the computing region (see Figure~\ref{G121314}). {It
should be noted that they are not necessarily open loops but may be connected to other places in the solar surface.}
L12 represents the bundles of magnetic field lines
rooted in N2, the brightening of G12 at the ending has good
agreement with L12 either in the view to the central region of the Sun or STEREO A/B side views. L13 are
also magnetic field lines rooted in N2. There are two bundles of field
lines that extend beyond the region. These two bundles of lines in L13 are
consistent with the EUV background G13 from all points of view.
The next group is L14 which displays a radial pattern rooted from P1.
These structures spread to the outside of the computing area and are in good
agreement with upper part of G14 in the central column in Figure~\ref{G121314}. {Some
open coronal loops in lower part of G14 in the central column are actually connected to lower magnetic pore region
outside AR 11158 as shown in Figure~\ref{fulldisk}(b), which are not included in the present magnetogram area employed
as boudnary condition. Therefore the corresponding lower part field lines of L14 at first follow the loop tendency but at higher altitude they bend back to deviate from the real coronal loops, which should be due to the ignorance of the boundary field
outside of the magnetogram area.}

\section{Conclusions} 
      \label{Conclusion}

The reconstructed topology configurations of the magnetic fields in NOAA 11158
is stereoscopically presented for the first time with comparison to observation from three points of view.
This allows us to understand this active region more comprehensively.
The calculated magnetic field lines replicate the observed EUV loop patterns well
from different views. These results demonstrate clearly that the DBIE method is effective when it is applied to
the actual photospheric magnetogram. {The GPU acceleration
makes DBIE tractable even if applied to large-scale domains. From the reconstructed coronal field structures,
we can estimate the altitude of EUV loop patterns which we found to be below 86 Mm, or~40\% of the
length of the magnetogram area.} They also match the actual EUV features as estimated from stereoscopic observations.
In this region the DBIE can achieve very high numerical accuracy. {In the present case for
$250\times200\times100$ internal grid points, the averaged angle between {\bf B} and {\bf J} is less than 4.5$^{\circ}$ and
the averaged relative flux error $\bar{g_i}=0.00067$.}

In the central sunspot cluster, the current density is very strong
along the filament in the PIL, and the current density distribution
on a vertical cross section was plotted (Sun et al., 2012). {Our results agree very well with
the situation that there are strong currents across the PIL and we found enlongate lower-lying twisting field lines co-spatial with the S-shaped filament along the PIL. However, we argue that one cannot simply attribute all the
EUV bright features along the PIL to manifestation of a filament although the filament could
be located there. Furthermore, we have obtained the electric current lines
three-dimensionally at higher altitude across the PIL in this region from three points of view.
According to their agreements with the bright EUV loop structures there,
we think that the features dominated by the strong currents should really exist
above the PIL.} Generally speaking, the region with strong currents
should contain a large amount of accumulated free energy and will eventually release
quickly in this region. {It is most possible that the extrapolated magnetic field lines
resembling the S-shaped filament
channel and the electric current lines agreeable with the bright EUV loops twistingly overlying the filament} may
be associated with the occurrence of the X2.2 flare.

It should be noted that while the line-of-sight (from the Earth direction) co-alignments between calculated field lines
and observed coronal loops agree with each other, the views from other sides may show
that they do not actually agree three-dimensionally and belong to other groups. This indicates
that co-alignment with line-of-sight images from the Earth direction alone may not provide the accurate coronal
configuration and the real three-dimensional information is vital in understanding the coronal
magnetic field structures and their associations with solar activities. {For the three stereoscopically reconstructed coronal loops, we quantitatively obtain the averaged misalignment angles of 16.6$^{\circ}$, 17.8$^{\circ}$, and 18.3$^{\circ}$ respectively, which are quite good with a factor of about two smaller than those given by other NLFFF models yielding overall misalignment angles of $24^{\circ} - 44^{\circ}$ (DeRosa et al. 2009), and at the same order as a forward-fitting model with reconstructed coronal loops as given conditions (Sandman and Aschwanden 2011).}

As a method different from others while they demonstrate similar computational capability, DBIE has the advantage that it just needs photospheric data as the boundary condition and allows to
evaluate the NLFFF field at every arbitrary point within the
domain from the boundary data instead of having to solve the entire domain. The
DBIE can be accelerated by parallel algorithm such as GPU techniques, which makes the DBIE
method be applied into larger boundary condition. The present study validates that the DBIE method is rigorous and practical.

In addition further acceleration could combine CUDA with MPI to
realize Muti-GPU parallelization, which will improve the
computational efficiency of DBIE method largely. As the first
images of Chinese Spectral Radioheliograph (CSRH, Yan et al., 2009) have been obtained, the
comparison between our extrapolation and the tomography
observation from CSRH will be carried out in the near future.

\begin{acks}
The authors would like to thank the referee for the helpful and valuable comments on this paper. Dr. Yingna Su is
acknowledged for improving the English of the manuscript and helpful comments. Mr. L. A. Selzer is acknowledged for improving the English of the manuscript as well. We thank the SDO and STEREO team for
providing the magnetic field data and EUV images used in this investigation.
We also wish to thank Dr. W.T. Thompson for his efficient support in the routine for correcting the STEREO data error.
This work is supported by NSFC Grants No. 11221063, 11273030, and
11211120147, MOST Grant No. 2011CB811401, and the National Major
Scientific Equipment R\&D Project ZDYZ2009-3. Part of experiments
were implemented on the ScGrid and GPU cluser of Supercomputing
Center, Computer Network Information Center of Chinese Academy of
Sciences.
\end{acks}


\end{article}

\end{document}